\documentclass[twocolumn,showpacs,preprintnumbers,superscriptaddress,amsmath,amssymb]{revtex4-1}
\usepackage{amssymb}		
\usepackage{amsmath}
\usepackage{graphicx}
\usepackage[normalem]{ulem}
\usepackage{multirow}
\usepackage{appendix}
\usepackage{float}
\usepackage{CJK}
\usepackage[usenames]{color}
\usepackage{bm}
\usepackage[colorlinks,linkcolor=blue,urlcolor=blue,citecolor=blue]{hyperref}
\usepackage{booktabs} 	     
\usepackage{leftidx}

\usepackage{soul,xcolor}
\setstcolor{red}
\newcommand\redsout{\bgroup\markoverwith{\textcolor{red}{\rule[0.5ex]{2pt}{0.4pt}}}\ULon}

\usepackage{tikz,xcolor,hyperref}

\definecolor{lime}{HTML}{A6CE39}
\DeclareRobustCommand{\orcidicon}{
	\begin{tikzpicture}
	\draw[lime, fill=lime] (0,0) 
	circle [radius=0.16] 
	node[white] {{\fontfamily{qag}\selectfont \tiny ID}};
	\draw[white, fill=white] (-0.0625,0.095) 
	circle [radius=0.007];
	\end{tikzpicture}
	\hspace{-2mm}
}
\foreach \x in {A, ..., Z}{%
	\expandafter\xdef\csname orcid\x\endcsname{\noexpand\href{https://orcid.org/\csname orcidauthor\x\endcsname}{\noexpand\orcidicon}}
}

\makeatletter

\newcommand{\Rmnum}[1]{\expandafter\@slowromancap\romannumeral #1@}

\makeatother
\begin{document}
\begin{CJK*} {UTF8} {gbsn}

\title{Production of $\Omega NN$ and $\Omega\Omega N$ in ultra-relativistic heavy-ion collisions}
\author{Liang Zhang(张良)}
\affiliation{Shanghai Institute of Applied Physics, Chinese Academy of Sciences, Shanghai 201800, China}
\affiliation{Key Laboratory of Nuclear Physics and Ion-beam Application (MOE), Institute of Modern Physics, Fudan University, Shanghai 200433, China}
\affiliation{School of Nuclear Sciences and Technology, University of Chinese Academy of Sciences, Beijing 100049, China}

\author{Song Zhang(张松)\orcidB{}}\thanks{Email: song\_zhang@fudan.edu.cn}
\affiliation{Key Laboratory of Nuclear Physics and Ion-beam Application (MOE), Institute of Modern Physics, Fudan University, Shanghai 200433, China}
\affiliation{Shanghai Research Center for Theoretical Nuclear Physics， NSFC and Fudan University, Shanghai 200438, China}

\author{Yu-Gang Ma(马余刚)\orcidB{}}\thanks{Email:  mayugang@fudan.edu.cn}
\affiliation{Key Laboratory of Nuclear Physics and Ion-beam Application (MOE), Institute of Modern Physics, Fudan University, Shanghai 200433, China}
\affiliation{Shanghai Research Center for Theoretical Nuclear Physics， NSFC and Fudan University, Shanghai 200438, China}

\date{\today}
\begin{abstract}
Even though lots of $\Lambda$-hypernuclei have been found and measured, multi-strangeness hypernuclei consisting of $\Omega$
are not yet discovered. The studies of multi-strangeness hypernuclei help us further understand the interaction between hyperons and nucleons. Recently 
the $\Omega N$ and $\Omega\Omega$ interactions as well as binding energies were calculated by the HAL-QCD's lattice Quantum Chromo-Dynamics (LQCD) simulations and  production rates of $\Omega$-dibaryon  
in Au + Au collisions at RHIC and Pb + Pb collisions at LHC energies were estimated by a coalescence model.
The present work discusses the production of more exotic triple-baryons including $\Omega$, namely $\Omega NN$ and $\Omega\Omega N$ as well as their decay channels. A variation method is used in calculations of  bound states and binding energy of $\Omega NN$ and $\Omega\Omega N$ with the potentials from the HAL-QCD's results. The productions of $\Omega NN$ and $\Omega\Omega N$ are predicted by using a blast-wave model plus coalescence model in ultra-relativistic heavy-ion collisions at $\sqrt{s_{NN}} = 200$ GeV and $2.76$ TeV. Furthermore, plots for baryon number dependent yields of different baryons ($N$ and $\Omega$), their dibaryons and hypernuclei are made and the production rate of a more exotic tetra-baryon ($\Omega\Omega NN$) is extrapolated.

\end{abstract}
\maketitle{}

\section{Introduction}

Hypernucleus consisting of hyperons and nucleons is  described by not only mass and charge but also hypercharge. Danysz and Pniewski first discovered the $^3_{\Lambda}H$ from cosmic rays in 1952~\cite{Danysz1953}. Since then more attention has been paid to hypernuleus research and many $\Lambda$-hypernuclei were discovered in cosmic rays as well as by accelerator beams~\cite{Davis2005,Gal}. Recently the observation of $\Xi^-$-$^{14}N$ was also  reported by the J-PARC laboratory~\cite{Hayakawa2020}. Nowadays, relativistic heavy-ion collisions can produce a large number of strange hyperons~\cite{Abelev2010,ChenRep,Feng,ZhangYF}, which provides a venue to discover the hypernucleus even anti-hypernucleus. 
The research on hypernuclei is becoming an important direction in heavy-ion collision experiments~\cite{Buyukcizmeci2020}. 
On the other hand, multi-quark exotic hadrons or hadronic molecules are also in current focus in particle and heavy-ion physics \cite{Rep1,Rep2,CPL1,CPL2,CPL3,Liao,PPNP}. 
The HAL-QCD Collaboration reported the most strangeness dibaryon candidates, $\Omega N$ and $\Omega\Omega$~\cite{Iritani2019,Gongyo2018} by the Lattice Quantum Chromo-Dynamics (LQCD) simulations. Based on their results, our previous work calculated the production of $\Omega\Omega$ and $\Omega N$ dibaryons and gave the yields of $\Omega$-dibaryon by the blast-wave model or A Multiphase Transport  (AMPT) model coupling with a coalescence model in relativistic heavy-ion collisions at $\sqrt{s_{NN}} = 200$ GeV and $2.76$ TeV \cite{Zhang2020}.

The attractive nature of the $\Omega N$ interaction leads to the possible existence of an $\Omega N$ dibaryon with strangeness = $-3$, spin = 2, and isospin = 1/2, which was first proposed in Ref.~\cite{Goldman}. Later on the HAL-QCD Collaboration calculated the $\Omega$-$N$ and $\Omega$-$\Omega$ interaction by the LQCD simulations near the physical point and the LQCD potentials are fitted by Gaussians and (Yukawa)$^2$. The results lead to the binding energy $B_{n\Omega} = 1.54$ MeV, $B_{p\Omega} = 2.46$ MeV and $B_{\Omega\Omega} = 1.6$ MeV~\cite{Iritani2019,Gongyo2018}. The STAR Collaboration made a first measurement of momentum  correlation functions of $p\Omega^-$ for Au + Au collisions at $\sqrt{s} = 200$ GeV \cite{NehaPLB} which indicates that the scattering length is positive for the proton-$\Omega$  interaction and favors the proton-$\Omega$ bound state hypothesis by comparing with the predictions based on the proton-$\Omega$ interaction extracted from (2 + 1)-flavor LQCD simulations \cite{Morita}.
Later on the ALICE collaboration measured the momentum correlation function of $p\Omega^-$ in pp collision at $\sqrt{s} = 13$ TeV \cite{Acharya2020} which supports the HAL-QCD result \cite{Iritani2019}. The potentials given by the HAL-QCD \cite{Iritani2019} are  also used to calculate the binding energy of $\Omega$-hypernuclei with $A$ = 3. Garcilazo and Valcarce \cite{Garcilazo2019} calculated the bound states of three-body $\Omega$-hypernuclei, namely $\Omega NN$ and $\Omega\Omega N$, by solving the Faddeev equations~\cite{FADDEEV1961} with the HAL-QCD potentials and obtained their binding energies ranging from 2 MeV to 20 MeV. 

In this work the productions of $\Omega NN$ and $\Omega\Omega N$ are calculated by a coalescence model in which the nucleon and hyperon phase space distributions are given by a blast-wave model~\cite{Retiere2004,PhysRevC.95.044905,PhysRevC.89.034918,Zhang2020}. The potentials from the LQCD are taken into account to obtain the relative wave functions and binding energies of $\Omega NN$ and $\Omega\Omega N$ by solving Schr\"odinger equation via a variation method. The estimation of the yields of $\Omega NN$ and $\Omega\Omega N$ will shed light on searching for $\Omega$-hypernuclei in experiment, such as at LHC-ALICE.

The calculation of production is introduced in Section II, which includes the brief introductions of the blast-wave  model and the coalescence model~\cite{Retiere2004,PhysRevC.95.044905,PhysRevC.89.034918,Zhang2020}, simplification of Wigner function as well as the variation calculation method of three-body bound state. It is  compared with the results from the Faddeev equations used by Garcilazo and Valcarce \cite{Garcilazo2019}. In Section III,  productions of $\Omega NN$ and $\Omega\Omega N$ are reported for Au + Au collisions at $\sqrt{s_{NN}} = 200$ GeV and Pb + Pb collisions at $\sqrt{s_{NN}} = 2.76$ TeV. The decay channels of $\Omega NN$ and $\Omega\Omega N$ are also discussed in this section.

\section{Method}
\subsection{Blast-wave model and coalescence model}

Cluster formation  in heavy-ion collision  can be realized  by the coalescence model ~\cite{Yan2006,PhysRevLett.84.4305,CHEN2003809,ZHANG2010224,PhysRevC.95.044905} or other methods like kinetic 
approaches \cite{Sun2021,Stau2021,NST_He}. 
In this work, a coalescence model constructed by the particle emission distribution and the Wigner density distribution 
is used to calculate few-body system production in heavy-ion collisions. The  multiplicity of three-constituent-cluster is given by,
 \begin{eqnarray}
  \label{eq_coalescence}
  \begin{aligned}
   N_{3 b} =g_{3} \int & \prod_{i=1}^3\left(d^{4} x_{i} S_{i}\left(x_{i}, p_{i}\right) \frac{d^{3} p_{i}}{E_{i}}\right) \times \\ 
   &\ \rho_{3}^{W}\left(x_{1}, x_{2}, x_{3}; p_{1}, p_{2},p_{3}\right),
  \end{aligned}
 \end{eqnarray}
where $\rho^W_3(x_1,x_2,x_3;p_1,p_2,p_3)$ is the Wigner density function which describes the coalescence probability, and $g_3 = (2S+1)/\left((2s_1+1)(2s_2+1)(2s_3+1)\right)$ is the coalescence statistical factor~\cite{Polleri1999,PhysRevC.98.054905-isospinNO,PhysRevC.92.064911-isospinNO,SUN2019132-isospinNO,Sun2020}, $S$ is the total spin for the three-body system and $s_i$ is the spin for each constituent particle. Table~\ref{tab_g3} lists the $g_3$ used in this paper for each $\Omega$-hypernucleus ($A$ = 3) and triton.
\begin{table}[]
\footnotesize
\caption{$g_3$ for $^3H$, $\Omega NN$ and $\Omega\Omega N$}
\label{tab_g3}
\begin{ruledtabular}
	\begin{tabular}{ccccccc}
   Nuclei   & $^3H$ & $\Omega pn$ & $\Omega nn$ & $\Omega pp$ & $\Omega\Omega n$ & $\Omega\Omega p$ \\
   $g_3$   & 1/4   & 3/8         & 1/4         & 1/4         & 1/16             & 1/16
	\end{tabular}
\end{ruledtabular}
\end{table}

In this work the particle emission distribution, $S_{i}\left(x_{i}, p_{i}\right)$, is given by the blast-wave model~\cite{Retiere2004,PhysRevC.95.044905,PhysRevC.89.034918,Zhang2020} which can describe the particle phase-space distribution in heavy-ion collisions. It assumes that in the rest frame the distribution of momenta is described by either a Bose or Fermi distribution of single particle and then the distribution is boosted into the center-of-mass frame of the total number of particles to describe the probability of finding a particle \cite{Cooper1974}. In heavy-ion collisions, the freeze-out time is considered following a Gaussian distribution \cite{Retiere2004,PhysRevC.95.044905,PhysRevC.89.034918,Zhang2020,NST_liu}. The blast-wave model is formalized as,
 \begin{equation}
  \label{eq_blw}
  \begin{aligned}
    S(x, p) d^{4} x = M_{T} \cosh \left(\eta_{s}-y_{p}\right) & f(x, p) \times\\
    & J(\tau) \tau d \tau d \eta_{s} r d r d \varphi_{s},
  \end{aligned}
 \end{equation}
where $M_T$ and $y_p$ are the transverse mass and the rapidity of a single particle, $r$ and $\varphi_{s}$ are the radius and azimuthal angle of  coordinate space, $\tau$ and $\eta_s$ are proper time and space pseudorapidity. $J(\tau)=\frac{1}{\Delta \tau \sqrt{2 \pi}} \exp \left[-\frac{\left(\tau-\tau_{0}\right)^{2}}{2(\Delta \tau)^{2}}\right]$ is the Gaussian distribution of freeze-out proper time, where $\tau_0$ and $\Delta \tau$ are the mean value and dispersion of this distribution. $f(x, p)=\frac{2 s+1}{(2 \pi)^{3}}\left[\exp \left(p^{\mu} u_{\mu} / T_{kin}\right) \pm 1\right]^{-1}$ is the Fermi or Bose distribution of a single particle boosted into the center-of-mass frame, where $s$ is the spin of the particle, $u_{\mu}$ is the four-velocity of a fluid element in the fireball of the particle source and $T_{kin}$ is the freeze-out temperature. The Lorentz invariant can be expressed as,
\begin{equation}
 \label{eq_pu}
 \begin{aligned}
  p^{\mu} u_{\mu}=M_{T} \cosh \rho_{\perp} &\cosh (\eta_s-y_p)-\\&p_{T} \sinh \rho_{\perp} \cos \left(\varphi_{p}-\varphi_{s}\right),
 \end{aligned}
\end{equation}
where $\varphi_{p}$ is the azimuthal angle in momentum space and $\rho_{\perp} $ is the transverse rapidity of fireball with a transverse radius $R_0$, defined as $\rho_{\perp} =v\rho_{0_{\perp}} \frac{r}{R_0}$. If the parameters ($\tau_0$, $\Delta \tau$, $\rho_{0_{\perp}}$, $R_0$ and $T_{kin}$) are fixed, the transverse momentum distribution is given as~\cite{Zhang2020}:
\begin{equation}
 \label{eq_pT_sep}
 \frac{d N}{2 \pi p_{T} d p_{T} d y_{p}}=\int S(x, p) d^{4} x.
\end{equation}

\subsection{Solving three-body bound state}

In order to obtain the Wigner function, bound state wave functions of $\Omega NN$ and $\Omega\Omega N$ need to be calculated. The non-relativistic Schr\"odinger equations of $\Omega NN$ and $\Omega\Omega N$'s bound state can be written as,
\begin{equation}
 \label{eq_Schr}
 \hat{H}\psi(\boldsymbol{x}_{1},\boldsymbol{x}_{2},\boldsymbol{x}_{3})=E_b\psi(\boldsymbol{x}_{1},\boldsymbol{x}_{2},\boldsymbol{x}_{3}),
\end{equation}
 \begin{equation}
  \label{eq_full_Schr}
  \begin{array}{c}
   \hat{H}=\displaystyle\sum_{i=1}^{3}-\frac{\nabla_{i}^{2}}{2 M_{i}} 
   +\sum_{j>i} V_{i j}\left(\boldsymbol{r}_{i j}\right),
  \end{array}
 \end{equation}
where $\displaystyle\psi(\boldsymbol{x}_{1},\boldsymbol{x}_{2},\boldsymbol{x}_{3}) = \sum_{i=1}^{3} \psi_{i}\left(\boldsymbol{x}_{i}\right)$ is total wave function of three-body system, $\psi_i(\boldsymbol{x}_i)$ and $\boldsymbol{x}_i$ are the wave function and coordinate of $i$-th particle, respectively, $\boldsymbol{r}_{ij}$ is relative coordinate between $i$-th and $j$-th particle defined as $\boldsymbol{r}_{ij} = \boldsymbol{x}_i-\boldsymbol{x}_j$. The potentials between $\Omega$-$N$ and  $\Omega$-$\Omega$ are the fit results from the  HAL-QCD simulation~\cite{Iritani2019,Gongyo2018}, and the $N$-$N$ potential is taken as the Malfliet-Tjon potential~\cite{Malfliet1969}: 
\begin{equation}
\label{eq_potential}
\begin{aligned}
&V_{NN}(r)=\sum_{i=1}^{2} C_{i} \frac{e^{-\mu_{i} r}}{r}, \\
&V_{N \Omega}(r)=b_{1} e^{-b_{2} r^{2}}+b_{3}\left(1-e^{-b_{4} r^{2}}\right)\left(\frac{e^{-M_{\pi} r}}{r}\right)^{2}, \\
&V_{\Omega \Omega}(r)=\sum_{i=1}^{3} C_{i} e^{-\left(r / d_{i}\right)^{2}},
\end{aligned}
\end{equation}
where $M_{\pi}$ is taken as $146$ MeV (near the physical mass $140$ MeV).  The parameters are listed in Table \ref{tab_potential_para}.

\begin{table}
\footnotesize
\caption{Parameters of potentials $V_{NN}(r)$~\cite{Malfliet1969}, $V_{N \Omega}(r)$, $V_{\Omega \Omega}(r)$~\cite{Iritani2019,Gongyo2018}}
\label{tab_potential_para}
\begin{ruledtabular}
	\begin{tabular}{lccccc}
		$V_{NN}(r)$ &				      & $C_1$ (MeV) & $C_2$ (MeV) & $\mu_1$ ($fm^{-1}$) & $\mu_2$ ($fm^{-1}$) \\
								& $^3S_1$ & -636.36		& 1460.47		  & 1.55								& 3.11								\\
								& $^1S_0$ & -521.74		& 1460.47		  & 1.55								& 3.11								\\
		\\ \hline\hline\\
		$V_{N \Omega}(r)$ &				  & $b_1$ (MeV) & $b_2$ ($fm^{-2}$) & $b_3$ (MeV$\cdot fm^{2}$) & $b_4$ ($fm^{-2}$) \\
								& $^5S_2$ & {-313.0~(5.3)}	& {81.7~(5.4)}		  	& {-252.0~(27.)}				& {0.85~(10)}							\\
		\\ \hline\hline\\
		$V_{\Omega \Omega}(r)$ & 				 & $C_1$ (MeV) & $C_2$ (MeV)    & $C_3$ (MeV)   &  \\
												 & $^1S_0$     & {914.0~(52)} 	    & {305.0~(44)}     & {-112.0~(13)}			 &  \\
												 &			   & $d_1\ (fm)$    & $d_2\ (fm)$   & $d_3\ (fm)$            &	\\
												 &			   & {0.143~(5)}       & {0.305~(29)}     & {0.949~(58)} 			 &	\\
	\end{tabular}
\end{ruledtabular}
\end{table}

There are many methods to solve this kind of three-body equations, such as the Faddeev equation~\cite{FADDEEV1961,Thompson2004,Kovalchuk2014} and the variation method. One kind of the variation methods is mainly based on the hyperspherical-harmonics (HH) method~\cite{Kievsky1993,Kievsky1994,Kievsky1997,Kievsky1997H}, 
in which the coordinates are transformed into center-of-mass frame by using the Jacobi transform,
{\setlength\abovedisplayskip{2pt}
\setlength\belowdisplayskip{2pt}
\begin{equation}
\label{eq_Jacobi_trans}
\begin{array}{c}
\left(\begin{array}{c}
\boldsymbol{R} \\
\boldsymbol{r}_{1} \\
\boldsymbol{r}_{2}
\end{array}\right)=J \cdot\left(\begin{array}{c}
\boldsymbol{x}_{1} \\
\boldsymbol{x}_{2} \\
\boldsymbol{x}_{3},
\end{array}\right) \\
\left(\begin{array}{l}
\boldsymbol{P} \\
\boldsymbol{q}_{1} \\
\boldsymbol{q}_{2}
\end{array}\right)=\left(J^{-1}\right)^{T} \cdot\left(\begin{array}{l}
\boldsymbol{p}_{1} \\
\boldsymbol{p}_{2} \\
\boldsymbol{p}_{3},
\end{array}\right)
\end{array}
\end{equation}
}
where $J$ is the Jacobi matrix, it reads 
\begin{widetext}
 \begin{equation}
  \label{eq_Jacobi_matrix}
  J = 
  \begin{pmatrix}
   \frac{M_{1}}{M_{tot}} & \frac{M_{2}}{M_{tot}} & \frac{M_{3}}{M_{tot}} \\[5pt]
   0 & -\sqrt{\frac{M_{2} M_{3}}{M_{23} Q}} & \sqrt{\frac{M_{2} M_{3}}{M_{23} Q}} \\[5pt]
   -\sqrt{\frac{M_{1}M_{23}}{M_{tot} Q}} & \sqrt{\frac{M_{2}^{2} M_{1}}{M_{tot}M_{23} Q}} & \sqrt{\frac{M_{3}^{2} M_{1}}{M_{tot} M_{23} Q}}
  \end{pmatrix},
 \end{equation}
\end{widetext}
where $M_i$ is the mass of $i$th particle, $M_{tot} = M_{1} + M_{2} + M_{3}$ is the total mass, $M_{23} = M_2 + M_3$ is the total mass of particles 2 and 3, $Q = \sqrt{(M_{1} M_{2} M_{3})/M_{tot}}$ is the reduced mass which normalizes the Jacobi matrix. For simplicity, the indexes of particles are chosen as symmetric as possible. In this article, the particles 2 and 3 prefer to be identical and particle 1 is different for a three-body nucleus. Sequentially the three-body Schr\"odinger equation separates into the center of mass motion (no effect on binding energy and relative wave function) and the relative motion~\cite{Chattopadhyay1996,Khan2012},
\begin{equation}
 \label{eq_Schr_six_dim}
 \widehat{T} \psi(\vec{r})+\sum_{j>i} V_{i j}\left(\boldsymbol{r}_{i j}\right) \psi(\vec{r}) = E_{b} \psi(\vec{r}), 
\end{equation}
where $\vec{r} = \left(\boldsymbol{r}_{1}, \boldsymbol{r}_{2}\right)=\left(\rho, \alpha, \theta_{1}, \phi_{1}, \theta_{2}, \phi_{2}\right)$ is defined in a six-dimensional hypersphere coordinate, $\rho = \sqrt{r_{1}^{2}+r_{2}^{2}}$ is the hyperradius, $\alpha = \arctan \left(r_{2}/r_{1}\right)$ is the hyperpolar angle which ranges from 0 to $\pi/2$~\cite{Chattopadhyay1996,Khan2012,He2015}, $\theta_i, \phi_i$ are the azimuth angles of $\boldsymbol{r}_i$, and the volume element is $d^6\vec{r} = \rho^5\sin^2{\alpha}\cos^2{\alpha}\sin{\theta_1}\sin{\theta_2}d\rho d\alpha d\theta_1 d\phi_1 d\theta_2 d\phi_2$. The momentum and angular momentum operators are defined as~\cite{Kievsky1993,Kievsky1994,Chattopadhyay1996,Khan2012,He2015},
\begin{equation}
 \label{eq_kinetic}
 \hat{T}=\frac{1}{2 Q}\left(-\frac{\partial^{2}}{\partial \rho^{2}}-\frac{5}{\rho} \frac{\partial}{\partial \rho}+\frac{\hat{L}^{2}}{\rho^{2}}\right),
\end{equation}
where 
\begin{equation}
 \label{eq_angular_momentum}
 \hat{L}^{2}=-\frac{\partial^{2}}{\partial \alpha^{2}}-4 \cot 2 \alpha \frac{\partial}{\partial \alpha}+\frac{\hat{l}_{1}^{2}}{\cos ^{2} \alpha}+\frac{\hat{l}_{2}^{2}}{\sin ^{2} \alpha}.
\end{equation}
The eigen function of $\hat{L}^2$ is a hyperspherical harmonic function~\cite{Kievsky1993,Kievsky1994,Kievsky1997,Kievsky1997H}:
\begin{widetext}
 \begin{eqnarray}
  \label{eq_hyper_spherical_harmonics}
  \begin{aligned}
   \mathcal{Y}_{K,\kappa}(\alpha,\theta_1,\phi_1,\theta_2,\phi_2)=N_{kl_{1}l_{2}} \cos(\alpha)^{l_1} \sin (\alpha)^{l_{2}} P_{k}^{l_2+1 / 2,\ l_1+1 / 2}(\cos(2 \alpha)) \times \\
   \left\{\left\{Y_{l_{1}}\left(\theta_{1}, \phi_{1}\right) Y_{l_{2}}\left(\theta_{2}, \phi_{2}\right)\right\}_{L}\left\{s_{i} s_{j k}\right\}_{S_{a}}\right\}_{JJ_z}\left\{t_{i} t_{j k}\right\}_{TT_z},
  \end{aligned}
 \end{eqnarray}
\end{widetext}
and 
\begin{align}
\label{eq_hyper_angular_eigen_value}
\hat{L}^2\mathcal{Y}_{K,\kappa}=(K+4)K\ \mathcal{Y}_{K,\kappa},
\end{align}
where $K = 2 k+l_1+l_2$ is the total hyperangular momentum number, $q$ is a nonnegative integer, $l_i$ and $m_i$ is the orbital angular momentum number of $\boldsymbol{r}_i$ direction, $\kappa$ represents the $L$-spin-isospin state defined as $\kappa = \{JJ_z(L(l_1l_2)S_a(s_is_{jk}))TT_z(t_it_{jk})\}$, $N_{kl_{1}l_{2}}$ is a normalization factor~\cite{Kovalchuk2014},
\begin{equation}
\label{eq_norm_factor}
N_{kl_{1}l_{2}} = \sqrt{\frac{2\ k!\ (K+2)\ (k+l_1+l_2+1)!}{\Gamma(k+l_1+3/2)\ \Gamma(k+l_2+3/2))}}
\end{equation}
and $P_{k}^{a,b}(x)$ is the Jacobi polynomial and $Y_{m}^l(\theta,\phi)$ is the Spherical Harmonic function. The orthogonal basis radial function can be chosen as 
\begin{widetext}
\begin{equation}
\label{eq_radial_basis}
u_n^{[\lambda]}(\rho)=\sqrt{\left(\frac{2\lambda}{n}\right)^3 \frac{(n-2)!}{2n\ (n+1)!}} e^{-\lambda\rho/n}\left(\frac{2\lambda \rho}{n}\right) L_{n-2}^3\left(\frac{2\lambda \rho}{n}\right)\rho^{-\frac{5}{2}}\ (n \geq 2), 
\end{equation}
\end{widetext}
in which $\lambda$ is a variation parameter, $n$ is the radial basis index, $L_a^b(\rho)$ is the associated Laguerre polynomial. Then the orthogonal basis function can be constructed as,
\begin{equation}
\label{eq_basis}
\left\langle\vec{r}\ |n,K,\kappa\right\rangle = u_n^{[\lambda]}(\rho)\mathcal{Y}_{K,\kappa}(\alpha,\theta_1,\phi_1,\theta_2,\phi_2).
\end{equation}

Then the relative motion Hamiltonian $\hat{H}$ can be expanded into matrix form $\left\langle n,K,\kappa\left|\hat{H}\right|n',K',\kappa'\right\rangle$. The following assumptions are taken to reduce the dimensions of the matrix: 1) assume that the nucleus is spherical by setting the total $L$ = 0, corresponding to the ground state; 
2) the $(I)J^{P}$ is fixed as the same as Garcilazo and  Valcarce  used \cite{Garcilazo2019};  3) if particle 2 and 3 are identical, the parity between them must be odd \cite{Kievsky1993}; 4)  $l_1$, $l_2\leq6$ is enough for required precision \cite{Kievsky1997}, and the number $n$ ranges from 2 to 11 and $k$ up to 45. The matrix elements have been calculated numerically by a Laguerre-Gauss quadrature for the integrals in the hyperradius $\rho$ and a Legendre-Gauss quadrature for the hyperangle $\alpha$
~\cite{HandbookofFunc}.

\begin{figure*}[htb]
\includegraphics[angle=0,scale=0.8]{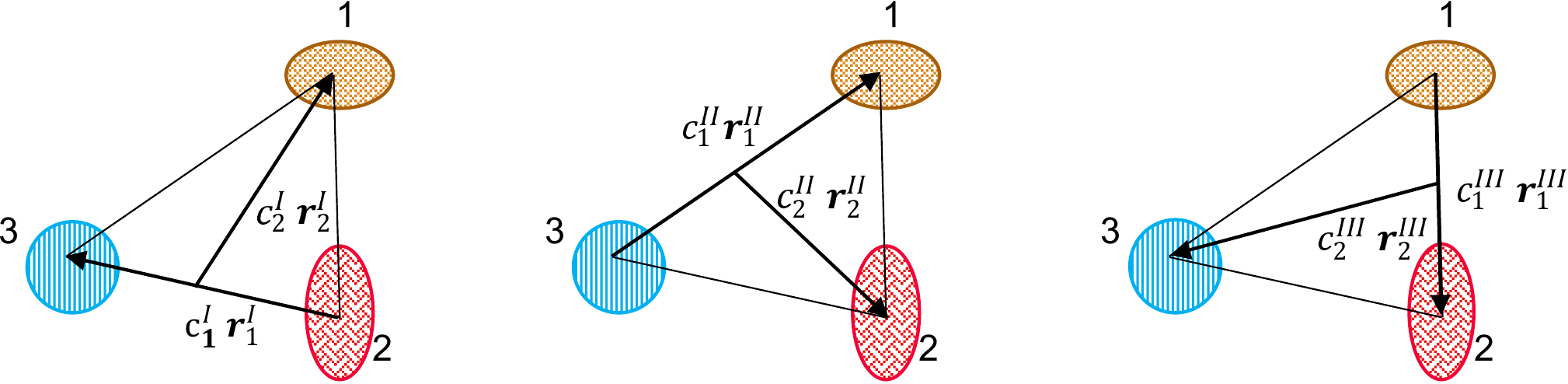}
\caption{Diagrams of three coordinate frames. From left to right, the coordinate is numbered as I, II and III, respectively, where $c^i_1 = \sqrt{\frac{M_{jk}Q}{M_j M_k}}$ and $c^i_2 = -\sqrt{\frac{M_{tot}Q}{M_i M_{jk}}}$ ($i,j,k = I(1), II(2), III(3)$ and $\epsilon_{ijk} = 1.$)}
\label{fig:fig-coor-tot}
\end{figure*}

But the elements of Hamiltonian matrix need six dimensional integral and the complex expressions of $\boldsymbol{r}_{12}$ and $\boldsymbol{r}_{31}$ for the hypersphere coordinate are based on the transforms of \eqref{eq_Jacobi_trans} and \eqref{eq_Jacobi_matrix}, where $\boldsymbol{r}_2 = -\sqrt{\frac{M_{2} M_{3}}{\left(M_{2} + M_{3}\right) Q}}\ \boldsymbol{r}_{23}$. In order to simplify the calculation, Raynal and Revai \cite{Raynal1970}
put forward the RR coefficient which is similar to Clebsch-Gordan coefficient. For example as shown in Fig.~\ref{fig:fig-coor-tot}, it is convenient to calculate $V_{23}(\boldsymbol{r}_{23})$ when the hypersphere is based on $\boldsymbol{r}_1^I$ and $\boldsymbol{r}_2^I$ in Coordinate I but hard to calculate $V_{12}(\boldsymbol{r}_{12})$ and $V_{31}(\boldsymbol{r}_{31})$ for the complex expressions of $\boldsymbol{r}_{12}$ and $\boldsymbol{r}_{31}$. By using RR coefficient, the hyperspherical harmonic function $\left|I;n,K,\kappa\right\rangle$, defined in the coordinate I, can be expanded by $\left|II(III);n,K,\kappa'\right\rangle$ in coordinate II (III),
\begin{widetext}
\begin{equation}
\label{eq_specific_basis}
\left|I;n,K,\kappa\right\rangle=\sum_{\kappa_k}\left\langle l^I_1l^I_2|l^{j}_1l^{j}_2\right\rangle_{K,L} \left\langle s_1s_{23};S|s_{1_j}s_{23_j};S\right\rangle \left\langle t_1t_{23};T|t_{1_j}t_{23_j};T\right\rangle \left|j;n,K,\kappa_j\right\rangle , 
\end{equation}
\end{widetext}
where $\left\langle l^I_1l^I_2|l^{j}_1l^{j}_2\right\rangle_{K,L}$ is the RR coefficient which requires that $K$ and $L$ are same in transformation, $\left\langle s_1s_{23};S|s_{1_j}s_{23_j};S\right\rangle$ and $\left\langle t_1t_{23};T|t_{1_j}t_{23_j};T\right\rangle$ are Clebsch-Gordan coefficients, $j$ represents the coordinate II or III, $1_j$ and $23_j$ represent the particle 2 (3) and the pair of particle 3 (1) and 1 (2) when $j = II (III)$. It is clear that the definition of $\rho$ is same in all coordinates, so the index $n$ does not need to change in the transform \eqref{eq_specific_basis}. After the transformation, $\boldsymbol{r}_{23_j}$ only relates to the $\rho$ and $\alpha_j$, which means the six dimensional integral is simplified into a double integral and a sum of $\kappa_j$.

After the calculation of Hamiltonian matrix, it is natural to calculate the minimum eigenvalue of the matrix as the binding energy $B[\lambda]$ of the three-body system and the corresponding eigenvector is the list of coefficients for the basis functions. And the binding energy $B[\lambda]$ requires $\delta B[\lambda]/\delta \lambda = 0$, which means that the binding energy is also the minimum point of variation parameter $\lambda$.

Garcilazo and Valcarce \cite{Garcilazo2016} solved three-body amplitudes by the Faddeev equations~\cite{FADDEEV1961} with considering the spin and isospin freedom. They assumed that three particles were in $S$-wave by which the spin-isospin state was   constructed and two-body amplitudes with the Legendre polynomials were expanded to solve the Faddeev equations.

Table \ref{tab_binging_E} shows the calculated binding energy of $^3H$, $_{\Lambda}^{3}H$ and $\Omega pn$ and the comparison with other theoretical results as well as experimental results. The potential between $N$ and $\Lambda$ used in $_{\Lambda}^{3}H$ binding energy calculation is YNG-ND interactions~\cite{Yamamoto1994,Hiyama1997} with $k_F = 0.84fm^{-1}$\cite{Hiyama2002}. It can be seen that this calculation of $\Omega pn$ is consistent with the results from Garcilazo and Valcarce's results ~\cite{Garcilazo2019}. {The error of $pn\Omega$ binding energy is estimated from the fitting errors of the $N-\Omega$ potential.}
The results of $^3H$ and $_{\Lambda}^{3}H$ are close to experimental results~\cite{Davis1986,STARCollaboration2020} and theoretical calculations as well~\cite{Egorov2021}. {Like $_{\Lambda}^{3}H$ consisting of spin $\frac{3}{2}$ and $\frac{1}{2}$, one is the ground state (spin $\frac{1}{2}$) and one is thought as a virtual state (spin $\frac{3}{2}$) near the $\Lambda d$ threshold \cite{Schafer2022}, $\Omega pn$ can also be mix of spin $\frac{5}{2}$, $\frac{3}{2}$ and $\frac{1}{2}$. According to the HAL QCD's calculation \cite{Gongyo2018}, the $^3S_1$ $\Omega N$ interaction is too weak to form a bound $\Omega N$ with spin 1. So the ratio of lower spin in $\Omega pn$ is small. In this paper $\Omega pn$ is considered as spin $\frac{5}{2}$.}

\begin{table}
\caption{The binding energies of $^3H$, $_{\Lambda}^{3}H$ and $\Omega pn$ calculated by a variation method. Results of this work are consistent with other's work and experimental results. 
The unit is in MeV}
\label{tab_binging_E}
\centering
\begin{center}
 \begin{ruledtabular}
	\begin{tabular}{ccccc}
	Nuclei			         & This work		&  Reference    &Experiment Result            \\
		$^3H$     			 & 8.69       &  8.4~\cite{Malfliet1969}	& 8.481~(Exp.) 		       \\
		\multirow{3}{*}{$_{\Lambda}^{3}H$}    & \multirow{3}{*}{2.68}      & \multirow{3}{*}{2.37~\cite{Egorov2021}}  & $2.35$~(Emulsion)~\cite{Davis1986} \\                         &                          &                                      & $2.63$~(STAR)~\cite{STARCollaboration2020}  \\                        &                          &                              & {$2.36$~(ALICE)~\cite{ALICECollaboration2019}}  \\
		$\Omega pn$          & {$22.0~(2.2)$}        &   21.3~\cite{Garcilazo2019} 			& / \\
	\end{tabular}
 \end{ruledtabular}
\end{center}
\end{table}

There is another method for few-body system interaction by which  the system is not deeply bound, called as the folding model \cite{Watanabe1958,Etminan2019}. The folding model assumes that nucleus is bound as a molecular state like dibaryon-baryon state. For the $^3S_1\ \Omega N$ interaction, it is not as strong as $^5S_2\ \Omega N$, the folding model can be applied in $\Omega nn$, $\Omega pp$, $\Omega\Omega n$ and $\Omega\Omega p$, 
which softens the $^5S_2\ \Omega N$ interaction. The model uses the free dibaryon wave function to average the potential between dibaryon and baryon,
\begin{small}
 \begin{equation}
 \label{eq_Folding}
  \begin{aligned}
   &U_{F}(\mathbf{R}_F) = \int d^3\mathbf{r}_d\ \psi_{d}^{*}\left(\mathbf{r}_d\right)\psi_{d}\left(\mathbf{r}_d\right)\times\\
   &\left[V_{12}\left(\mathbf{R}_F-\frac{M_{3}\ \mathbf{r}_d}{M_{2}+M_3} \right)+V_{13}\left(\mathbf{R}_F+\frac{M_{2}\ \mathbf{r}_d}{M_{2}+M_3} \right)\right], 
  \end{aligned}
 \end{equation}
\end{small}
where $\psi_d$ is the dibaryon wave function which is consisted of particle 2 and 3, $U_F(\mathbf{R}_F)$ is average potential and the $\mathbf{R}_F$ is relative coordinate between the dibaryon and baryon. This method also simplifies the three-body bound state into two two-body bound states (dibaryon and dibaryon-baryon). The total wave function is $\Psi(\mathbf{R}_F,\mathbf{r}_d) = \psi_d(\mathbf{r})\ \psi_{mole}(\mathbf{R}_F)$ and total binding energy is $E = E_d + E_{mole}$, where $\psi_{mole}(\boldsymbol{R}_F)$ is the molecular state wave function calculated with the average potential $U_F(\boldsymbol{R}_F)$ and $E_{mole}$ is the binding energy of molecular state. The binding energies of $\Omega nn$, $\Omega pp$, $\Omega\Omega n$ and $\Omega\Omega p$ are calculated by the folding model and {their   errors  are estimated from the fitting error of $N-\Omega$ and $\Omega-\Omega$ potential. The results are listed} in Table \ref{tab_folding_binging_E}.
It can be found that different combinations of dibaryon in three-body systems result in different binding energies which are corresponding to different decay channels and  will be discussed later.
\begin{table}
\centering
\caption{Comparison of {binding energies} between folding model calculation (this work) and the work of Garcilazo and Valcarce (Ref.~\cite{Garcilazo2019}).  It seems that $\Omega NN$ and $\Omega\Omega N$ whose spin is smaller than 5/2 is weakly bound to the third baryon}
\label{tab_folding_binging_E}
 \begin{ruledtabular}
  \begin{tabular}{cccc}
   Nuclei													 & dibaryon-baryon 	 & this work & reference~\cite{Garcilazo2019} \\
   $\Omega nn$     									 & $\Omega n+n$      & {$2.81 ~(1.26)$}      & 2.35      \\
   $\Omega pp$     									 & $\Omega p+p$      & {$4.02 ~(1.26)$}      & 3.04      \\
   \multirow{2}{*}{$\Omega\Omega n$} & $\Omega n+\Omega$ & {$6.77 ~(3.17)$}      & \multirow{2}{*}{5.1}       \\
        														 & $\Omega\Omega +n$ & {$4.83~(3.13)$}  		 &           \\
   \multirow{2}{*}{$\Omega\Omega p$} & $\Omega p+\Omega$ & {$10.2 ~(3.2)$}      & \multirow{2}{*}{6.5}     \\
        														 & $\Omega\Omega +p$ & {$6.22~(3.32)$} 		 &           \\
 \end{tabular}
 \end{ruledtabular}
\end{table}

\subsection{Wigner function}

\begin{figure}[thb]
    \centering
    \includegraphics[angle=0,scale=0.45]{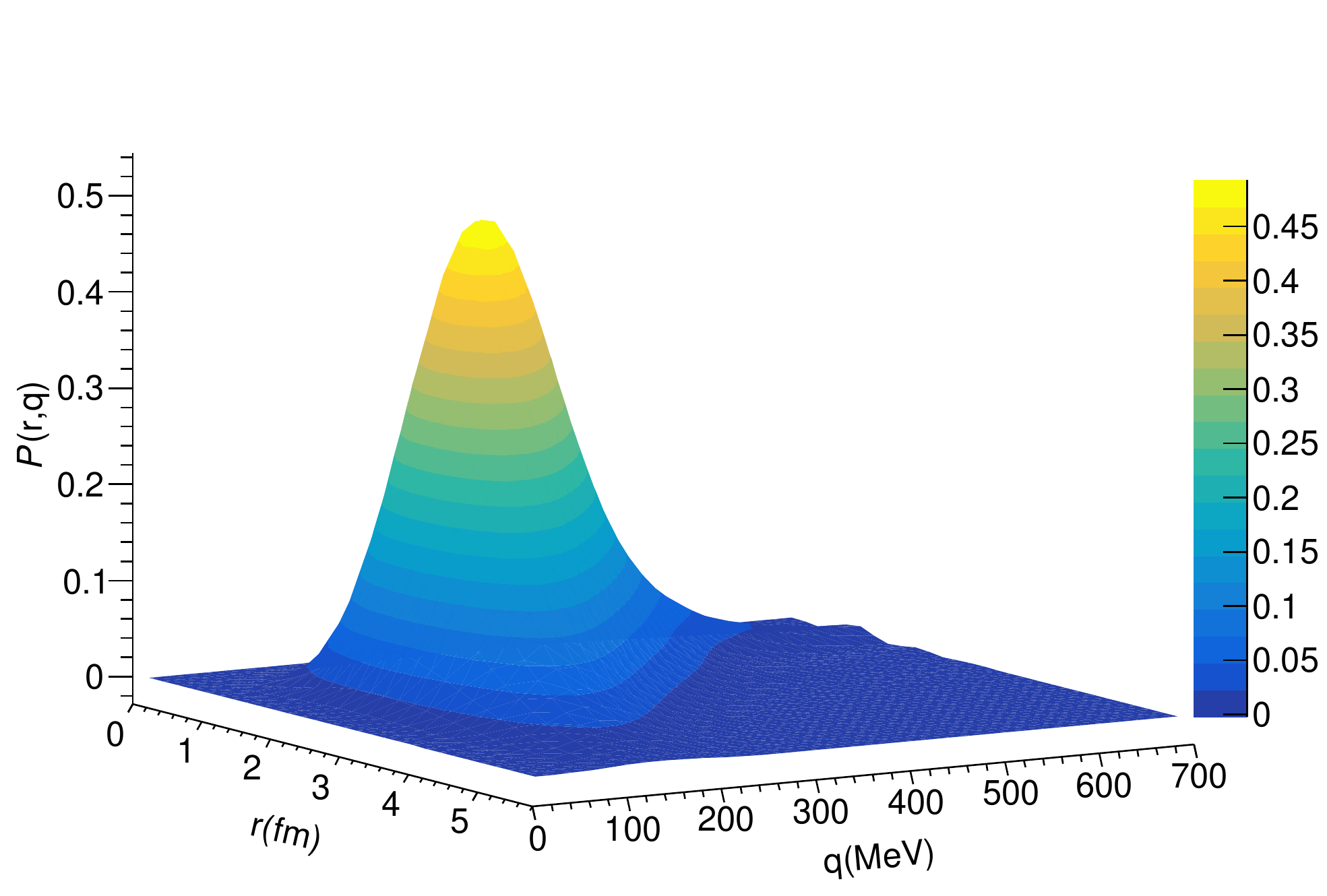}
    \caption{Wigner probability $P(r,q)$ of $pn\Omega$, which represents the probability to find the $pn\Omega$ ground bound state at binding energy 21.7 MeV at six-dimensional hyperspherical radius $r$ and at six-dimensional hyperspherical momentum $q$}
    \label{fig:fig_wig_show}
\end{figure}

The Wigner function introduced in Eq.~(\ref{eq_coalescence}) is written as~\cite{CHEN2003809,ZHANG2010224,PhysRevC.95.044905},
\begin{equation}
\label{eq_Wigner_function}
 \begin{aligned}
  \rho^{W}(\vec{r}, \vec{q})=\int \psi\left(\vec{r}+\frac{\vec{R}}{2}\right) &\psi^{*}\left(\vec{r}-\frac{\vec{R}}{2}\right) \times\\
  & \exp (-i \vec{q} \cdot \vec{R}) d^6 \vec{R},
 \end{aligned}
\end{equation}
where $\vec{r} = (\boldsymbol{r}_1, \boldsymbol{r}_2)$, $\vec{q} = (\boldsymbol{q}_1, \boldsymbol{q}_2)$ are the relative coordinate and momentum, and $\psi(\vec{x})$ is the relative wave function. For the three-body system it is expressed in six dimensions, the Wigner function will be 12 dimensions, which is impossible to draw a picture and hardly calculated. After performing  the calculation of eigenvector of Hamiltonian matrix, the major contribution of total wave function comes from a few bases which contribute more than $94\%$ to total amplitude for the parameters of them are large (larger than 0.08). 
{With considering the fitting errors of potential, the total relative errors of
such simplified wave functions are about $10\%$, So this kind of simplification retains most information of origin wave function. If the selected bases are only radial related, the total wave function can be simplified as the sum of these bases with weights of their parameters. And then the simplified wave function is only radial related.}
The Wigner function can be simplified as,
\begin{small}
 \begin{equation}
 \label{eq_specific_Wigner}
  \begin{aligned}
   \rho_{3}^{W}(r, q, \theta)=&\displaystyle\int \psi\left(\sqrt{r^{2}+\frac{R^{2}}{4}+r R \cos \theta_{1}}\right) \times \\
   &\ \ \ \ \ \psi^{*}\left(\sqrt{r^{2}+\frac{R^{2}}{4}-r R \cos \theta_{1}}\right) \times \\
   &\ \ \ \ \ \ \ \exp \left(-i q R \cos \theta \cos \theta_{1}\right) \times \\
   &\ \ \ \ \ \ \ \ \exp \left(-i q R \sin \theta \sin \theta_{1} \cos \theta_{2}\right) \times \\
   &\ \ \ \ \ \ \ \ \ \ 2 \pi^{2} R^{5} \sin^{4} \theta_{1} \sin^{3} \theta_{2} d R d \theta_{1} d \theta_{2}.\\
  \end{aligned}
 \end{equation}
\end{small}
A Laguerre-Gauss quadrature is applied for the integrals of hyperradius $R$ and $\theta_1,\ \theta_2$ is integrated by a Legendre-Gauss quadrature~\cite{HandbookofFunc}. The coordinate is defined in a six-dimensional spherical coordinate as $\vec{R}=\left(R, \theta_{1}, \theta_{2}, \theta_{3}, \theta_{4}, \theta_{5}\right)$, which can be transformed into the six-dimensional Cartesian coordinate:
\begin{equation}
 \begin{aligned}
  \vec{R}=&(\boldsymbol{R}_1,\boldsymbol{R}_2)=(R_{1x},R_{1y},R_{1z},R_{2x},R_{2y},R_{2z})=\\
  &\ (R\cos{\theta_1},R\sin{\theta_1}\cos{\theta_2},R\sin{\theta_1}\sin{\theta_2}\cos{\theta_3},\\
  &\ \ R\sin{\theta_1}\sin{\theta_2}\sin{\theta_3}\cos{\theta_4},\\
  &\ \ R\sin{\theta_1}\sin{\theta_2}\sin{\theta_3}\sin{\theta_4}\cos{\theta_5},\\
  &\ \ R\sin{\theta_1}\sin{\theta_2}\sin{\theta_3}\sin{\theta_4}\sin{\theta_5}),   
 \end{aligned}
\end{equation}
and $0\leq\theta_i\leq\pi$ ($i=1,2,3,4$), $0\leq\theta_5\leq2\pi$, the volume element is $d^6\vec{R}  = R^5dR\prod_{i=1}^5\sin^{5-i}{\theta_i}d\theta_i$. The $\vec{r}$ in Eq.~\eqref{eq_specific_Wigner} is set at $\left(r,0,0,0,0,0\right)$, the $\vec{q}$ is set at $(q,\theta,0,0,0,0)$. By integrating out the angle, the probability to find the $pn\Omega$ ground bound state can be obtained at six-dimensional hyperspherical radius $r$ and at six-dimensional hyperspherical momentum $q$~\cite{He2015},
\begin{equation}
\label{eq_prob}
P(r,q) = \frac{1}{24\pi}r^5q^5\int_0^{\pi}\rho_3^W(r,q,\theta)\sin^4\theta d\theta,
\end{equation}
which is shown in Fig.~\ref{fig:fig_wig_show}. The Wigner probability is similar to a Gaussian distribution with tails in  both coordinate and momentum space. The most probable position in the coordinate-momentum phase space is located at $(r,q)\sim(2\text{ fm}, 200\text{ MeV})$.
And the normalization of the probability,
\begin{equation}
\label{eq_prob_norm}
\int^{\infty}_0P(r,q)drdq = 1.
\end{equation}

If the wave function relates to not only $\rho$ but also $\alpha$, in other word, the wave function relates to both $r_1$ and $r_2$ which are defined in Fig.~\ref{fig:fig-coor-tot}. Wigner transformation is more complex. $\psi(\rho,\alpha)=\left.\left\langle\rho\alpha\right|\psi\right\rangle$ can be simplified into $\sum_{n_1,n_2} \left.\left\langle r_1 r_2\right| n_1 n_2\right\rangle\left\langle n_1 n_2\left|\psi\right\rangle\right.$. $\left\langle r_i\left|n_i\right\rangle\right.$ is a 3-dimension radial orthogonal basis which is the same as \eqref{eq_radial_basis} but the last term is $r_i^{-1/2}$ with the same variation parameter $\lambda$ for different $n_i$. Here $n_i$ ranges from 2 to 26 with $\lambda = 10000$. By this way, Wigner transformation can be rewritten as:
\begin{small}
 \begin{equation}
  \label{eq:sixdim-wigner}
  \begin{aligned}
   \rho^{W}(\vec{r}, \vec{q})&=\int  d^6 \vec{R} \left\langle\psi\left|\vec{r}-\frac{\vec{R}}{2}\right\rangle\right. \left.\left\langle\vec{r}+\frac{\vec{R}}{2}\right|\psi\right\rangle e^{ -i \vec{q} \cdot \vec{R} }\\
   &=\sum_{n_1,n_2,n'_1,n'_2}\left.\left\langle\psi\right| n_1 n_2\right\rangle\left\langle n'_1 n'_2\left|\psi\right\rangle\right.\times\\
   &\prod_{i=1,2}\int d^3 \mathbf{R_i} \left\langle n_i\left|\mathbf{r}_i-\frac{\mathbf{R}_i}{2}\right\rangle\right.\left.\left\langle\mathbf{r}_i+\frac{\mathbf{R}_i}{2}\right| n'_i\right\rangle e^{ -i \mathbf{q}_i \cdot \mathbf{R}_i }.
  \end{aligned}
 \end{equation}
\end{small}
A complex Wigner transformation is simplified by a series of three-dimension Wigner transform.

For the folding model, $\rho_{3}^{W} = \rho_{di}^W\times \rho_{di-b}^W$, where $\rho_{di}^W$ is the Wigner density function for dibaryon and $\rho_{di-b}^W$ is the Wigner density function for the pair of dibaryon and third baryon. Both of these two Wigner density functions can be calculated as did in our previous work~\cite{Zhang2020} for two-body systems.

{The main errors of Wigner function are from the errors of wave functions.  From the relationship between Wigner function and the wave function, the errors of Wigner function are estimated to be about $20\%$.}

\section{Result and discussion}

\begin{table}[t]
\centering
\caption{The blast-wave model parameters for proton ($\Omega$) in Au + Au collisions at $\sqrt{s_{NN}} = 200$ GeV \cite{Sun2015}, which is fitted from the RHIC data \cite{Adler2004,TheSTARCollaboration2009} as well as in Pb+Pb collisions at $\sqrt{s_{NN}} = 2.76$ TeV \cite{Zhang2020} fitted from the ALICE data \cite{Abelev2013,Adam2016,Abelev2014}
}
\label{tab_coalsece_para}
\footnotesize
\begin{ruledtabular}
 \begin{tabular}{cccccc}
          & T (MeV) & $\rho_0$   & $R_0$ (fm) & $\tau_0$ (fm$/c$) & $\Delta \tau$ (fm$/c$) \\
 200 GeV  & 111.6   & 0.98 (0.9) & 15.6     & 10.55           & 3.5                  \\
 2.76 TeV & 122     & 1.2 (1.07) & 19.7     & 15.5            & 1                    \\
 \end{tabular}
\end{ruledtabular}
\end{table}

\begin{figure*}[thb]
\centering
\includegraphics[angle=0,scale=0.9]{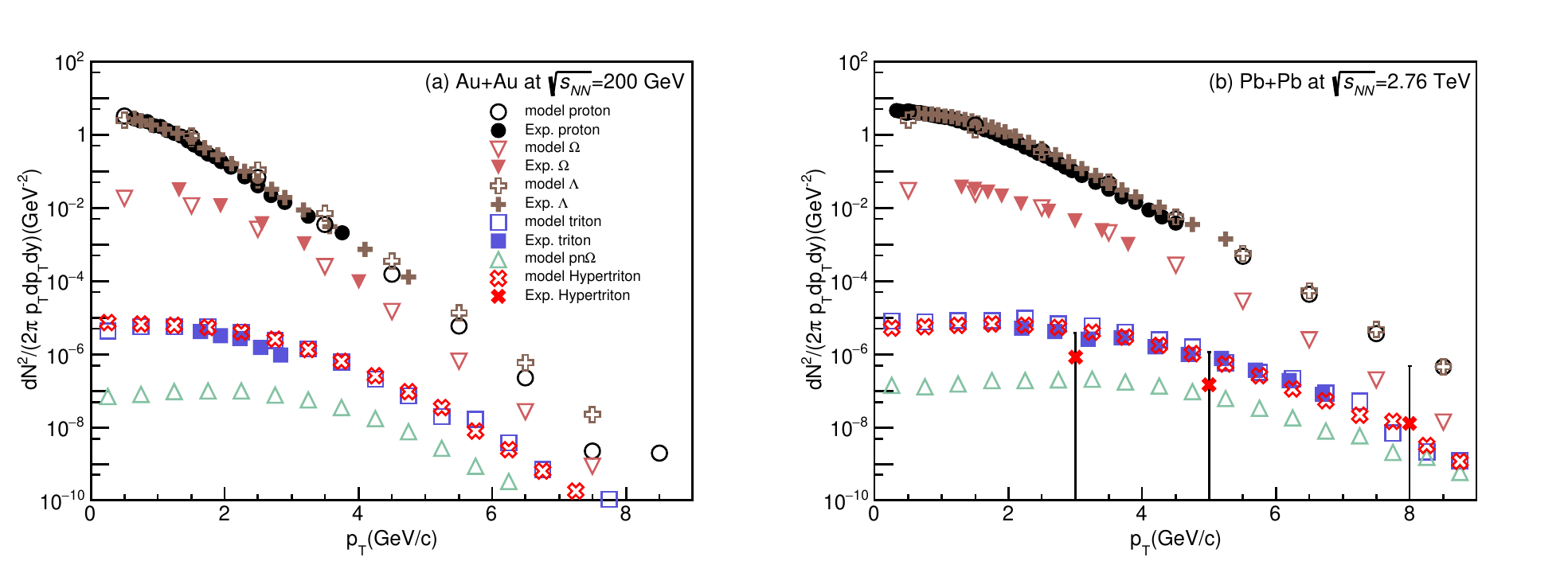}
\caption{Transverse momentum $p_T$ spectra of $p,\ \Omega,\ ^3_{\Lambda}H,\ \text{ and } \Omega pn$ in Au + Au collisions at $\sqrt{s_{NN}}  = 200$ GeV (a) and Pb + Pb collisions at $\sqrt{s_{NN}} = 2.76$ TeV (b). The open markers for $p$, $\Lambda$ and $\Omega$ directly fit to the experiments, and the open makers for triton, hypertriton and $\Omega pn$ are the results of BLWC model. Full makers are the data from the RHIC~\cite{Adler2004,Adams2007,Zhang2021} and  ALICE~\cite{Abelev2013,Abelev2014,Adam2016,ALICECollaboration2015}
}
\label{fig_pT_spectra}
\end{figure*}

In blast-wave model, the parameters ($\tau_0$, $\Delta \tau$, $\rho_{0_{\perp}}$, $R_0$ and $T_{kin}$) are fitted with experimental transverse momentum spectra of proton and $\Omega$ by Eq. \eqref{eq_pT_sep} and adjusted with the results of triton for different collisions, as shown in Fig.~\ref{fig_pT_spectra}. Table \ref{tab_coalsece_para} listed the parameters used in this work. 

\begin{figure*}[thb]
\centering
\includegraphics[angle=0,scale=0.9]{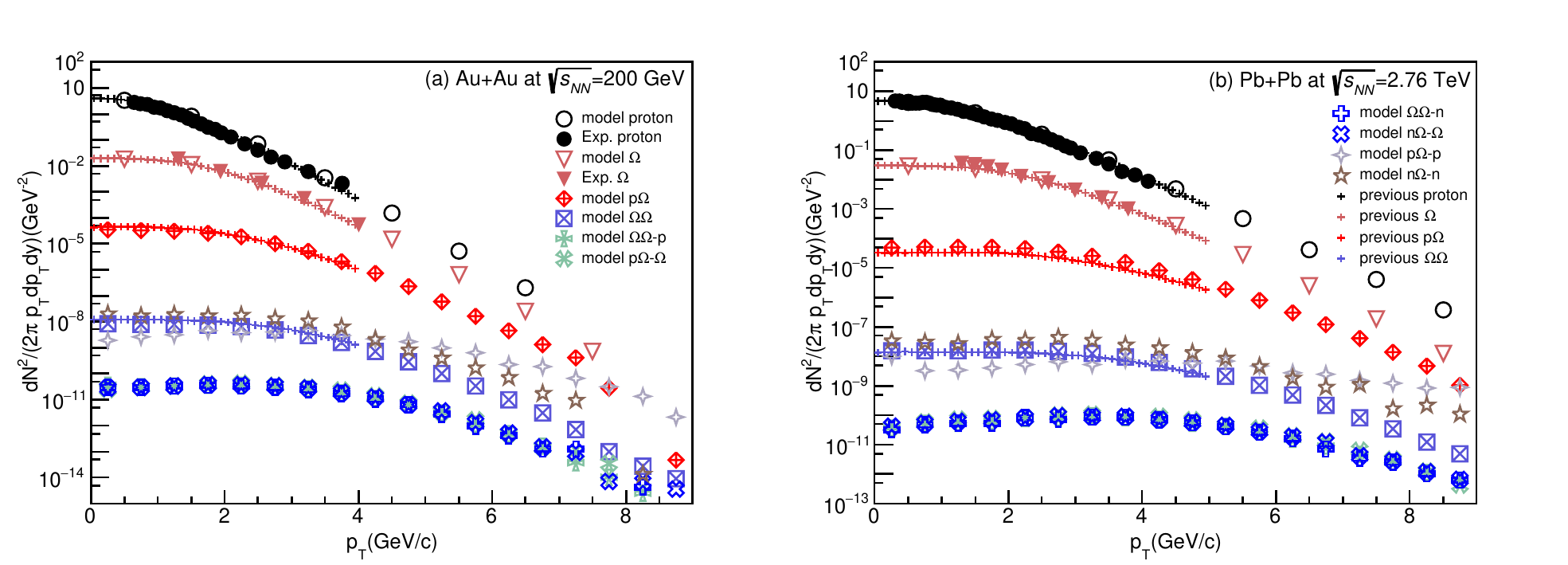}
\caption{Transverse momentum $p_T$ spectra of proton calculated by the folding model, the open makers for $\Omega nn,\ \Omega pp,\ \Omega\Omega n \text{ and } \Omega\Omega p$ are the results of the BLWC model and the folding model. The full makers are the data from the RHIC \cite{Adler2004,Adams2007,Zhang2021} and  ALICE \cite{Abelev2013,Abelev2014,Adam2016}. The lines are calculated by our previous work \cite{Zhang2020}}
\label{fig_pT_spectra_folding}
\end{figure*}

The transverse momentum spectra of $\Omega pn$ is calculated by using the blast-wave model coupled with coalescence model (BLWC) as Eq. \eqref{eq_coalescence} and shown in Fig.~\ref{fig_pT_spectra} (a) for Au+Au collisions at $\displaystyle\sqrt{s_{NN}} = 200$ GeV and Fig.~\ref{fig_pT_spectra} (b) for Pb + Pb collisions at $\displaystyle\sqrt{s_{NN}} = 2.76$ TeV. The results of $\Omega nn,\ \Omega pp,\ \Omega\Omega n \text{ and } \Omega\Omega p$ with the relative wave function from the folding model are shown in Fig.~\ref{fig_pT_spectra_folding}. The $p_T$ spectra of $p\Omega$ and $\Omega\Omega$ from our previous work~\cite{Zhang2020} as well as this work are also presented in Fig.~\ref{fig_pT_spectra_folding}. The $p_T$ spectra of $n\Omega$ is not shown here because it is almost as same as $p\Omega$.

\begin{figure}[thb]
\centering
\includegraphics[angle=0,scale=0.4]{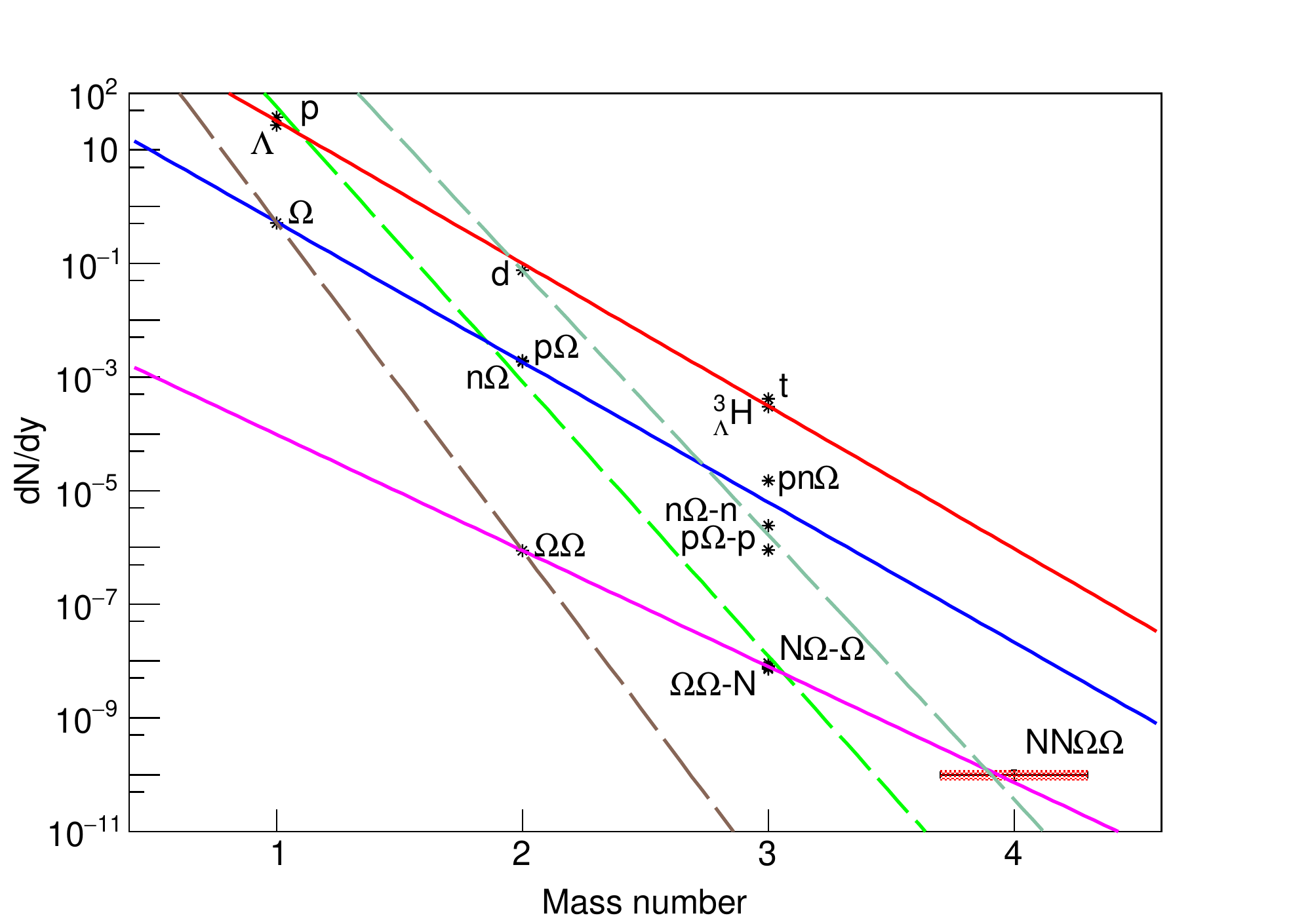}
\caption{The exponential decay relation of $dN/dy$ versus the constituent mass number ($A$) for Pb + Pb collisions at 2.76 TeV. There are basically two-class production chains, namely the first class: $N\rightarrow d \rightarrow t ~(^3_{\Lambda}H)$ (red), $\Omega\rightarrow N\Omega\rightarrow NN\Omega$ (blue), $\Omega\Omega\rightarrow N\Omega\Omega$ (pink), and the second class: $N \rightarrow N\Omega \rightarrow N\Omega\Omega$ (green), $\Omega \rightarrow \Omega\Omega$ (brown) and $d \rightarrow N\Omega N \rightarrow NN\Omega\Omega$ (light green) chains. These lines show the relation $dN/dy\sim\exp(-bA)$, where $ b$ = 5.78 (red), 5.68 (blue), and 4.70 (pink) for the first class, and  11.1 (green), 13.3 (brown) and 10.7 (light green)
}
\label{fig_dNdy_A}
\end{figure}

To further investigate the productions of $\Omega$-dibaryons and hypernuclei, the $p_T$ integrated yields $dN/dy$ at midrapidity are given in Table~\ref{tab_dN/dy} and ~\ref{tab_dN/dy_folding}. The predicted results show  $N\Omega$ $\sim\times 10^{-4}$~\cite{Zhang2020}, $\Omega\Omega$ $\sim\times 10^{-7}$, $\Omega NN$ $\sim\times 10^{-7}$ and $N\Omega\Omega$ $\sim\times 10^{-9}$. { The uncertainties of the integrated yields are directly from the Wigner functions, whose relative errors are about $20\%$. So the relative errors of yields are considered as $20\%$. Though the uncertainties from the blast-wave parameters are also important, which have been discussed by other model work \cite{PhysRevC.89.034918}, it will not be discussed in this paper.} And while, the corresponding values in Pb + Pb collisions at 2.76 TeV are larger than those in Au + Au collisions at 200 GeV. With the growing of constituents number $A$ such as $\Omega \rightarrow N\Omega \rightarrow NN\Omega$ and $N \rightarrow N\Omega \rightarrow \Omega\Omega N$, the production rates appear to follow the exponential function $\exp(-bA)$, here $b$ is the so-called reduction factor~\cite{SHAH20166,PhysRevC.85.064912,STARantiAlpha}, as shown in Fig.~\ref{fig_dNdy_A} for Pb + Pb collisions at 2.76 TeV. This $A$-dependent trend is similar to that for light nuclei of $p \rightarrow d \rightarrow t ~(^3_{\Lambda}H)$ in Fig.~\ref{fig_dNdy_A}. However, it can be seen that $n\Omega$-$n$ ($p\Omega$-$p$) slightly deviate from the trend in $\Omega \rightarrow N\Omega \rightarrow NN\Omega$. Keep in mind that the treatment of interaction is slightly different between $pn\Omega$ and dibaryon-baryon via the folding method, which results in the slight deviation.  In general,  we have two classes for these production chains. One is for  $N \rightarrow d \rightarrow t~(^3_{\Lambda}H)$, $\Omega \rightarrow N\Omega \rightarrow NN\Omega$ and $\Omega\Omega \rightarrow N\Omega\Omega$ (solid lines), they are almost parallel with the increase of $N$ constituent number. Another is for $N \rightarrow N\Omega \rightarrow N\Omega\Omega$, $\Omega \rightarrow \Omega\Omega$ and $d \rightarrow NN\Omega$ chains (dash lines), they are almost parallel  with the increase of $\Omega$ number. Obviously much larger reduction factor $b$ for the second class than the first class, indicating that much less yield for adding one more $\Omega$ than one more nucleon. The different reduction factor $b$ results from the different interactions between $N-\Omega$ and $\Omega-\Omega$ as well as the difference of productions of $N$ and $\Omega$. Inspired by this, the production of hypernuclei $N_nH_m$ ($N$ for nucleons and $H$ for one kind of hyperons) can be estimated by the intersection of $N_iH_m$ and $N_nH_j$ chains ($i(j)$ is smaller than $n(m)$). Even if there is one point on the chain of $N_nH_j$, the reduction factor $b$ of this chain is similar to the chain of $H_j$ or other chains whose $b$ is known in the same class with $N_nH_j$. From figure~\ref{fig_dNdy_A}, the prediction of the $NN\Omega\Omega$ production is about $10^{-10}$. It implies that the production of hypernuclei is sensitive to the interaction among the constituents in the coalescence framework and then the systematic measurement of hypernuclei can shed light on the production mechanism and the baryon interaction.

\begin{table}[thb]
\centering
\caption{$dN/dy$ for $\Omega$-dibaryons and hypernuclei 
at midrapidity. The values of $dN/dy$ for $\Omega$-dibaryons 
are taken from Ref.~\cite{Zhang2020}}
\label{tab_dN/dy}
\begin{ruledtabular}
 \begin{tabular}{ccccc}
          & $n\Omega$           & $p\Omega$           & $\Omega\Omega$      & $\Omega pn$         \\
 200 GeV  & $7.51\times10^{-4}$ & $7.39\times10^{-4}$ & $3.1\times10^{-7}$ & $3.56\times10^{-6}$ \\
 2.76 TeV & $1.31\times10^{-3}$ & $1.27\times10^{-3}$ & $7.9\times10^{-7}$ & $7.36\times10^{-6}$ \\
 \end{tabular}
\end{ruledtabular}
\end{table}

\begin{table}[b]
 \centering
 \caption{$dN/dy$ for $\Omega$-dibaryon and hypernuclei
 at midrapidity calculated by the folding model. The values of $dN/dy$ for $\Omega$-dibaryons are re-calculated in this work}
 \label{tab_dN/dy_folding}
 \begin{ruledtabular}
  \begin{tabular}{ccccc}
            & $p\Omega$          & $\Omega\Omega$       & $n\Omega-n$ 		  & $p\Omega-p$         \\
   200 GeV  & $6.84\times10^{-4}$ & $2.51\times10^{-7}$ & $6.95\times10^{-7}$ & $2.78\times10^{-7}$ \\
   2.76 TeV & $1.95\times10^{-3}$ & $8.87\times10^{-6}$ & $2.37\times10^{-6}$ & $9.00\times10^{-7}$ \\
   \\ \hline\hline \\
            & $n\Omega-\Omega$    & $\Omega\Omega-n$    & $p\Omega-\Omega$    & $\Omega\Omega-p$    \\
   200 GeV  & $1.87\times10^{-9}$ & $1.69\times10^{-9}$ & $2.03\times10^{-9}$ & $1.76\times10^{-9}$ \\
   2.76 TeV & $8.11\times10^{-9}$ & $7.29\times10^{-9}$ & $9.12\times10^{-9}$ & $7.62\times10^{-9}$ \\
  \end{tabular}
 \end{ruledtabular}
\end{table}

$\Omega NN$ can weak decay through an $\Omega$ decay, which  decays into $\Lambda$-hypernuclei ($A$ = 3) or $\Xi$-hypernuclei ($A$ = 3), note that $\Xi$-hypernuclei ($A$ = 3) can not be formed according to the HAL-QCD's results but might be formed under the ESC08c potential~\cite{Hiyama2020}.  $\Omega NN$ can also strong decay into $\Lambda\Xi N$ or $\Sigma\Xi N$ which is based on the interaction $\Omega N-\Lambda\Xi$ and $\Omega N-\Sigma\Xi$ reported by the HAL-QCD~\cite{Sekihara2018}. As for $\Omega \Omega N$, it can decay into $N\Omega\Lambda$ or $N\Omega\Xi$ and mesons from the weak decay of $\Omega$. It can also decay into $\Lambda\Xi\Omega$ or $\Sigma\Xi\Omega$ by strong interaction. All here mentioned three-baryon group, such as $\Lambda\Xi N$ and $N\Omega\Lambda$, may not be bounded. 

From Fig.~\ref{fig_pT_spectra_folding}, it is hard to figure out the difference between $\Omega\Omega$-$N$ and $\Omega N$-$\Omega$. Although the $p_T$ spectra of $\Omega\Omega$-$N$ and $\Omega N$-$\Omega$ are almost the same, the strong decay channels are different in the folding model. For $\Omega N$-$\Omega$, it would decay into a $\Lambda \Xi$ or $\Sigma\Xi$ through the $\Omega N-\Lambda\Xi$ or $\Omega N-\Sigma\Xi$ channel and an $\Omega$, while the $\Omega\Omega$-$N$ can hardly decay into $\Lambda \Xi \Omega$ or $\Sigma\Xi\Omega$, since the $N$ and $\Omega$ are not bound directly in this folding model.

\section{Summary}

The three-body bound state problem can be solved through a variation method coupled with an eigenvalue problem. For weakly-bounded triple-particle system, the folding model is applied. The $N$-$\Omega$ and $\Omega$-$\Omega$ potentials used in this work are fitted from the lattice QCD's simulation near the physical point,  which was reported by HAL-QCD collaboration. In coalescence model, the phase-space information of nucleons and $\Omega$ are generated by the blast-wave model and the particles are coalesced into $\Omega NN$ and $\Omega\Omega N$ by using the Wigner density function from the simplified three-body wave function. 
The production of $NN\Omega$ is about $10^{-7}$ and $N\Omega\Omega$ is about $10^{-9}$. There are also $A$-dependent trends similar with that for $p\rightarrow d\rightarrow t~(^3_{\Lambda}H)$. The production rates follow the exponential function $\exp(-bA)$. With adding different baryon, the reduction factor $b$ is different. Due to different factor $b$, two classes of hypernuclei chains will intersect at certain points where the production rate of new hypernucleus could be estimated.
And the decay modes of $\Omega NN$ and $\Omega\Omega N$ are briefly discussed in order to search for such exotic triple-baryons 
(hypernuclei) in future experiments, which could provide a method to understand the $YN$ and $YY$ interactions for multi-strangeness hadrons. The systematic measurements of hypernuclei can definitely shed light on the production mechanism and baryon interactions.

\begin{acknowledgements}

This work was supported in part by the National Natural Science Foundation of China under contract Nos. 11875066, 11890710, 11890714, 11925502, 11961141003, and 12147101, National Key R\&D Program of China under Grant No. 2018YFE0104600 and 2016YFE0100900, the Strategic Priority Research Program of CAS under Grant No. XDB34000000, and Guangdong Major Project of Basic and Applied Basic Research No. 2020B0301030008.

\end{acknowledgements}



\end{CJK*}	

\bibliography{reference}

\end{document}